\documentstyle[epsfig,onecolumn]{mn}

\def\kpc{{\rm\thinspace kpc}}

\def\kmps{\hbox{$\km\s^{-1}\,$}}

\def\keV{{\rm\thinspace keV}}

\def\km{{\rm\thinspace km}}
\def\s{{\rm\thinspace s}}

\def\keV{{\rm\thinspace keV}}

\def\erg{{\rm\thinspace erg}}
\def\ergps{\hbox{$\erg\s^{-1}\,$}}

\def\pcmcu{\hbox{${\rm cm}^{-3}\,$}}

\title[Viscous intracluster medium]{
Buoyant radio-lobes in a viscous intracluster medium
}

\author[C.~S.~Reynolds et al.]{
\parbox{14cm}{Christopher~S.~Reynolds$^1$,
Barry~McKernan$^1$,
Andrew~C.~Fabian$^2$,
James~M.~Stone$^3$, and
John~C.~Vernaleo$^1$}\\
$^1$Dept.\ of Astronomy, University of Maryland, College Park, MD 20742, USA.\\
$^2$Institute of Astronomy, Madingley Road, Cambridge, CB3~OHA\\
$^3$Department of Astrophysical Sciences, Peyton Hall, Princeton, NJ08544, 
USA\\
}

\date{In press}
\pagerange{\pageref{firstpage}--\pageref{lastpage}}
\pubyear{2004}

\begin{document}
\label{firstpage}

\maketitle

\begin{abstract}
  Ideal hydrodynamic models of the intracluster medium (ICM) in the
  core regions of galaxy clusters fail to explain both the observed
  temperature structure of this gas, and the observed morphology of
  radio-galaxy/ICM interactions.  It has recently been suggested that,
  even in the presence of reasonable magnetic fields, thermal
  conduction in the ICM may be crucial for reproducing the temperature
  floor seen in many systems.  If this is indeed correct, it raises
  the possibility that other transport processes may be important.
  With this motivation, we present a numerical investigation of the
  buoyant evolution of AGN-blown cavities in ICM that has a
  non-negligible shear viscosity.  We use the ZEUS-MP code to follow
  the 3-dimensional evolution of an initially static, hot bubble in a
  $\beta$-model ICM atmosphere with varying degrees of shear
  viscosity.  With no explicit viscosity, it is found that the
  combined action of Rayleigh-Taylor and Kelvin-Helmholtz
  instabilities rapidly shred the ICM cavity and one does not
  reproduce the intact and detached ``ghost cavities'' observed in
  systems such as Perseus-A.  On the other hand, even a modest level
  of shear viscosity (corresponding to approximately 25\% of the
  Spitzer value) can be important in quenching the fluid instabilities
  and maintaining the integrity of the bubble.  In particular, we show
  that the morphology of the NW ghost cavity found in Perseus-A can be
  reproduced, as can the flow pattern inferred from the morphology of
  H$\alpha$ filaments.  Finally, we discuss the possible relevance of
  ICM viscosity to the fact that many of the active ICM cavities
  (i.e., those currently associated with active radio-lobes) are not
  bounded by strong shocks, the so-called ``shock problem''.
\end{abstract}


\section{Introduction and observational background}

Recent years have seen a growing realization that the cores of rich
galaxy clusters are complex and dynamic environments.  In particular,
it is becoming clear that the radio-loud active galactic nuclei (AGN)
often hosted by the cD galaxies in rich clusters can have a major
influence on the hydrodynamics and thermodynamics of the core regions
of the intracluster medium (ICM).  As we discuss below, the rich
datasets coming from the {\it Chandra X-ray Observatory} and {\it
XMM-Newton} now demand theoretical models that go beyond the simple
picture of jet-blown ``bubbles'' rising in a static ICM described by
ideal hydrodynamics.  Bulk ICM motions (including turbulence),
magnetohydrodynamics (MHD), thermal conductivity, and viscosity may
all be relevant to data that are currently being taken.

Even before the launches of {\it Chandra} and {\it XMM-Newton}, {\it
  Einstein} and {\it ROSAT} studies revealed prominent
radio-galaxy/cluster interactions in three clusters; Perseus-A
(B\"ohringer et al. 1993; Heinz, Reynolds \& Begelman 1998), Virgo-A
(Feigelson et al. 1987; B\"oringer et al.  1995) and Cygnus-A
(Carilli, Perley \& Harris 1994; Harris, Carilli \& Perley 1994).
{\it ROSAT} showed Perseus-A and Cygnus-A to possess intracluster
medium (ICM) cavities coincident with the prominent radio-lobes in
these two sources, suggesting supersonic inflation of a bubble in the
ICM by the jetted AGN (Clarke, Harris \& Carilli 1997).  Virgo-A, on
the other hand, showed a cooler (and soft X-ray brighter) region of
ICM associated with the outer eastern ``ear'' seen in low-frequency
radio maps of Virgo-A (Owen, Eilek \& Kassim 2000).  It was first
suggested by B\"ohringer et al.  (1995) that this phenomenon might be
caused by lower entropy gas from the cluster center being dragged
upwards in the ICM atmosphere by a buoyantly rising radio-lobe.  

More recent observations by {\it Chandra} and {\it XMM-Newton} show
that radio-galaxy induced ICM substructure is surprisingly ubiquitous
and complex.  The basic results described above, i.e., the existence
of ICM cavities associated with active radio-lobes and the presence of
cool material that appears to lie in the wake of old, buoyantly rising
radio-lobes, have been confirmed in numerous systems (e.g., Hydra-A;
McNamara et al. 2000, Abell~2052; Blanton et al. 2001, Virgo-A; Young,
Wilson \& Mundell 2002, Perseus-A; Fabian et al. 2000, 2003a).
However, these new data have raised several mysteries and are
increasingly at odds with simple models for radio-galaxy/ICM
interactions.  Firstly, in any ideal hydrodynamic model, the cavities
must be inflated supersonically or else they would be destroyed by
Rayleigh-Taylor (RT) instabilities faster than they are inflated.
Curiously, the strong and hot ICM shocks that one expects to find
around the active cavities are notably absent; instead, many ICM
cavities are surrounded by ICM shells that are {\it cooler} than the
ambient ICM.  We shall refer to this as the ``shock problem''.
Secondly, ICM cavities that are not associated with any obvious
radio-lobe (``ghost cavities'') have been discovered.  Examples are
found in the Perseus cluster (Fabian et al. 2003a), Abell~2597
(McNamara et al. 2001), and Abell~4059 (Heinz et al. 2002, Choi et al.
2004).  In some cases, ghost cavities are coincident with regions of
low-frequency (74\,MHz) radio emission supporting the hypothesis that
they correspond to old radio-lobes from previous and now extinct AGN
outbursts (Fabian et al. 2002a).  Interestingly, as we will explicitly
demonstrate in this paper, ideal hydrodynamic models fail to reproduce
the observed morphology of at least some ghost cavities.  Finally,
most clusters have been found to possess a ``temperature-floor'' in
the sense that the radiative cooling of the ICM appears not to proceed
below temperatures of $kT\sim 1-2\keV$ (Tamura et al. 2001; Peterson
et al.  2001).  Again, no clear explanation for this fact is provided
by an ideal hydrodynamic model.  Br\"uggen \& Kaiser (2002) suggest
that stirring of the ICM core by a central radio galaxy may be
responsible for the temperature floor; however, such a scenario only
postpones rather than prevents radiative cooling, and is hard to
reconcile with the strong metallicity gradients observed in the cores
of some clusters (David et al. 2001, Matsushita et al. 2002; Sanders
\& Fabian 2002).  This has led several authors (Narayan \& Medvedev
2001; Fabian et al.  2002b; Voigt et al. 2002; Kim \& Narayan 2003) to
resurrect the notion that thermal conduction may be important in
determining the thermodynamics of ICM cores.

Faced with the multiple failures of simple hydrodynamic models for
cluster cores, we must carefully examine the other physical processes
that might be relevant.  Both the dynamical and thermodynamical
effects of magnetic fields are largely unexplored and are the subject
of on-going large-scale simulations.  However, the qualitative success
of models including thermal conduction in solving the temperature
floor problem begs a study of other transport processes and, in
particular, the effects of shear viscosity on the dynamics of a
radio-galaxy/ICM interaction.

\begin{figure}
\centerline{
\psfig{figure=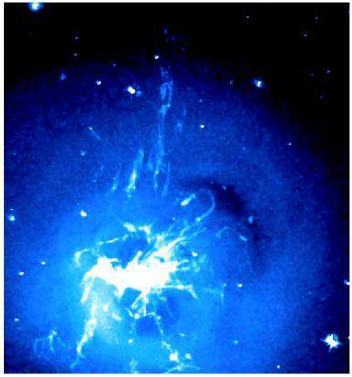,width=0.42\textwidth}
}
\caption{X-ray image of the core regions of the Perseus cluster 
  from the {\it Chandra X-ray Observatory} (blue; from Fabian et al.
  2003a) overlaid with the H$\alpha$ image from the WIYN telescope
  (Fabian et al. 2003b; Conselice et al. 2001).  The flattened X-ray
  cavity is clearly visible in the central regions of this image.
  Furthermore, the H$\alpha$ emitting filaments display well-defined
  arcs, suggestive of a vortex-like flow pattern in the region behind
  the buoyant cavity.  This image is 4.33\,arcmin (96\,kpc) on a side
  and is oriented such that North is upwards.}
\label{fig:perseus}
\end{figure}

Perseus-A and the core of the Perseus cluster continues to be one of
the best studied examples of a rich cluster core and a complex
radio-galaxy/cluster interaction.  The most detailed X-ray
investigation to date, based on a deep (200\,ks) observation by {\it
  Chandra}/ACIS-S, has been presented by Fabian et al. (2003a).  These
authors report the discovery of wave-like disturbances in the ICM on
spatial scales of $\sim 50\kpc$, approximately twice the spatial
scales of the most obvious ghost cavity to the north-west of
Perseus-A.  They discuss a scenario in which viscous dissipation of
these disturbances may act as a significant heat source for the ICM
core.  It is shown that, provided viscosity operates reasonably close
to its ideal unmagnetized value, it is possible for viscous
dissipation of radio-galaxy induced disturbances to balance radiative
cooling of the ICM.  This suggestion has been supported by recent
simulation work by Ruszkowski, Br\"uggen \& Begelman (2003).
Circumstantial evidence for the presence of significant ICM viscosity
is also provided by an examination of the morphology of H$\alpha$
filaments.  Several of the filaments appear to trace well-defined arcs
in the region below the ghost cavity (Fig.~\ref{fig:perseus}; also see
Fabian 2003b).  This argues against the presence of strong turbulence
in the ICM core, possibly resulting from the action of viscosity.  If
we make the stronger presumption that the H$\alpha$ filaments follow
streamlines in the ICM, the morphology of the filaments suggests the
existence of a vortex ring within the ICM just below the NW ghost
cavity.

With this background and motivation, this paper presents hydrodynamic
simulations of the buoyant evolution of an AGN-blown cavity in a
viscous ICM.  In order to bring clarity to the discussion of such a
complex system, this paper deals with the focused question of how ICM
viscosity effects the observed morphology and associated flow patterns
of old (ghost) cavities.  Detailed investigations of the effects of
viscosity on the growth of active cavities and the thermodynamic state
of the ICM will be addresses in future work.  Section~2 reviews the
importance of viscosity in typical clusters, and touches upon the
robustness of viscosity in the presence of magnetic fields.  Section~3
describes the basic set-up of our simulations, as well as our results
on the morphology and flow patterns.  Section~4 discusses some of the
limitations of this work, and possible implications of ICM viscosity.
Finally, conclusions are presented in Section~5.

\section{The importance of viscosity in the ICM}
\label{sec:viscosity}

Before discussing simulations of viscous systems, we shall address in
brief the general issue of viscosity in the ICM.  Initially suppose
that the ICM can be described as a thermal fully-ionized plasma which
is unmagnetized.  The relevant coefficient of viscosity is given by
Braginskii (1958) and Spitzer (1962) as
\begin{equation}
\mu_{\rm um}=2.2\times 10^{-15}\frac{T^{5/2}}{\ln\Lambda}\,{\rm g}\,{\rm cm}^{-1}\,{\rm s}^{-1},
\label{eq:mu}
\end{equation}
where $T$ is the temperature of the plasma measured in Kelvin and
$\ln\Lambda$ is a Coulomb integral.  Scaling to a temperature of
$kT=5T_5\keV$ and using $\ln\Lambda=30$ gives $\mu=1.88\times
10^3T_5\,{\rm g}\,{\rm cm}^{-1}\,{\rm s}^{-1}$.

It is customary to measure the importance of viscosity in a fluid of
density $\rho$ through the Reynolds number, $Re=UL/\nu$, where
$U$ and $L$ are characteristic velocities and length scales of the
system and $\nu=\mu/\rho$.  Of course, in a complicated system such as
a radio-galaxy/ICM interaction, there is no unique velocity and length
scale and so it is not possible to define a universal Reynolds numbers
that characterizes the system.  However, provided one uses some
consistent choice for $U$ and $L$, the Reynolds number is still useful
as a means to parameterize the relative importance of viscosity
between different AGN/ICM systems.  It also allows a means of matching
the viscosity imposed in simulations with that expected in real
systems.

We choose the maximum dimension of the bubble as our characteristic
length scale, and half of the adiabatic sound speed (i.e. a typical
buoyancy-induced rise velocity) as our characteristic velocity.
Scaled to the NW ghost cavity of the Perseus cluster (with number
density $n\approx 0.03\pcmcu$, and $kT\approx 5\keV$ which gives
$c_s^2=780\kmps$), gives a Reynolds number of
\begin{equation}
Re=62\left(\frac{U}{390\kmps}\right)\left(\frac{L}{20\kpc}\right)\left(\frac{kT}{5\keV}\right)\left(\frac{n}{0.03\pcmcu}\right)\left(\frac{\mu}{\mu_{\rm um}}\right)^{-1}.
\end{equation}
We note that this Reynolds number is almost an order of magnitude
smaller than the fiducial value of $Re=400$ quoted by Robinson et
al. (2003), primarily due to our higher (and more realistic) fiducial
temperature.  Thus, restating the conclusion of Fabian et al. (2003a),
viscosity may be relevant to the evolution of an AGN
induced bubble in the ICM.

The major uncertainty is the effect that magnetic fields have on the
macroscopic viscosity.  The case of a uniform field is readily
analyzed (Spitzer 1962).  The proton gyro-radius corresponding to any
non-negligible magnetic field is very small, leading to extremely
efficient suppression of the local coefficient of viscosity
perpendicular to the magnetic field; for typical ICM conditions, the
perpendicular coefficient of viscosity is suppressed by the enormous
factor of $\sim 10^{23}$ (Spitzer 1962).  However, the effective
macroscopic viscosity in the case of a realistic magnetic field
configuration (which is almost certainly tangled, and may be chaotic)
is an open question.  A similar issue has recently been addressed in
the context of thermal conduction.  In that case, the local thermal
conductivity is also suppressed by a very large factor perpendicular
to the field.  However, the exponential divergence of neighbouring
field lines in a chaotic field structure results in an effective
thermal conductivity, $\kappa$, that is suppressed below the
unmagnetized value , $\kappa_{\rm um}$ by only a factor of
$10^{-2}-0.2$ (depending on the spectrum of fluctuations in the field
structure; Narayan \& Medvedev 2001).  While the tensorial nature of
the viscous stress tensor prevents a precise mapping of the two
problems, similar arguments may apply and we might expect the
effective coefficient of viscosity to be suppressed below the
unmagnetized value by some factor ranging from $10^{-2}$ to unity.

Clearly, the effective thermal conductivity and viscosity
characterizing the ICM is still very much an open theoretical
question, due to uncertainties in both the basic physics of transport
processes in hot plasmas as well as the magnetic field structure
present in the ICM.  To make progress we must {\it assume} that
certain conditions exist, compute the consequences for
radio-galaxy/ICM interactions, and compare with the recent
observations.  This is the motivation for the rest of this paper.
While the evidence for non-negligible ICM viscosity is still
circumstantial, we show that the action of such a viscosity allows the
morphology of ghost cavities (in particular, the NW ghost cavity in
Perseus-A) to be reproduced, and may be an important mechanism for
stabilizing the ICM core against radiative losses.

\section{Viscous hydrodynamic simulations}

\subsection{Basic setup of the simulations}

Our goal is to study the evolution of a buoyant radio lobe in the ICM
of a galaxy cluster, including the effects of kinematic viscosity.
Following the approach of Bruggen \& Kaiser (2001), we simulate only
the phase of the evolution after the radio-loud AGN has inflated a
low-density bubble which has expanded to achieve pressure equilibrium
with the ICM.  At some point during this process, the AGN activity is
assumed to cease.  Thus, our simulation follows the evolution of a
low-density bubble, initially in pressure equilibrium, placed in the
central regions of an ICM atmosphere.  

Even though real radio lobes will have rapid and complex internal
flows (induced by the AGN jet during their inflation), we shall assume
that both the ICM and the bubble interior are initially static.  We
stress that this is an important simplification in our models and must
be kept in mind when interpreting the results.  For example, our
calculations will not be meaningful for addressing the issue of shocks
and sound waves driven into the ICM by the radio-galaxy, since these
phenomena are almost certainly dominated by the jet-driven inflation
phase of the bubble which we are not modeling.  Furthermore, the
precise growth rates of the RT and Kelvin-Helmholtz (KH) will be
influenced by the internal motions within the bubble left over from
its inflation phase.  However, our set-up allows a qualitative
investigation of the hydrodynamics of buoyantly rising ICM bubbles.

Our detailed set-up is as follows.  The undisturbed ICM atmosphere is
given a density profile described by
$\rho(r)=\rho_0\left[1+(r/r_0)^2\right]^{-0.75}$, where we choose
units of mass and length such that $\rho_0=1$ and $r_0=1$.  This
atmosphere is assumed to be initially static and isothermal, with an
adiabatic sound speed of $c_s=1$.  The gravitational potential,
$\Phi$, is assumed to be dominated by dark matter and, hence, is
assumed to be fixed throughout the simulation.  This gravitational
potential is determined by the condition that the initial ICM
atmosphere is in hydrostatic equilibrium, $\nabla p=-\rho\nabla\Phi$,
where $p$ is the pressure.  In this ICM atmosphere, we carve out a
spherical bubble with density $\rho_{\rm bub}=0.01$ and radius $r_{\rm
  bub}=1/4$.  This bubble is displaced from the center of the ICM
atmosphere by $\Delta r=1/4$ (i.e., the boundary of the bubble touches
the center of the ICM atmosphere). The initial pressure of the bubble
is set equal to the initial ICM pressure at that radius (i.e., this is
a hot, low-density bubble is in local pressure equilibrium with the
ICM).  We note that this is a very simplified form for the cluster
density profile and gravitational potential.  However, we feel that a
more sophisticated cluster profile (e.g., using a Navarro, Frenk \&
White [1997] profile, and including the potential of the cD galaxy)
would be unwarranted for the qualitative nature of the current
investigation.

This initial state is evolved using the equations of 3-dimensional
viscous hydrodynamics,
\begin{eqnarray}
\frac{D\rho}{Dt}+\rho\nabla\cdot {\bf v}&=&0\\
\rho\frac{D{\bf v}}{Dt}&=&-\nabla p - \rho\nabla\Phi - \nabla\cdot {\bf \Pi}\\
\rho\frac{D}{Dt}\left(\frac{\epsilon}{\rho}\right)&=&-p\nabla\cdot {\bf v}-{\bf \Pi}:\nabla {\bf v}
\end{eqnarray} 
where ${\bf v}$ is the velocity field, $\epsilon$ is the internal
energy density, and ${\bf \Pi}$ is the viscous stress tensor,
\begin{equation}
\Pi_{ik}=\mu \left(\frac{\partial v_i}{\partial x_k} + \frac{\partial v_k}{\partial x_i} - \frac{2}{3}\delta_{ik}\frac{\partial v_j}{\partial x_j}\right)
\end{equation}
Note that we only include ``p\,dV'' and viscous dissipation terms in
the energy equation, eqn.~4.  In particular, we neglect the radiative
cooling which is unlikely to be relevant for the dynamical timescale
processes modeled here.  These equations are solved using the ZEUS-MP
MHD code (Stone \& Norman 1992a, 1992b), operating in Cartesian
co-ordinates $(x,y,z)$. ZEUS is a fixed-grid, time-explicit Eulerian
code which uses an artificial viscosity to handle shocks.  When
operated using van Leer advection, as in this work, it is formally of
second order spatial accuracy.  We model one half of the ICM
atmosphere using a simulation volume of a cubic box of length $2r_0$
on a side.  The cluster center ($r=0$) is placed in the center of one
face of this cube, and reflecting boundary conditions are imposed on
this face in order to account for the unmodelled half of the
atmosphere.  Outflow boundary conditions are imposed on all other
faces.

The effects of viscosity are introduced into the basic ZEUS-MP code by
adding explicit source terms to the momentum and energy equations
(eqns. 4 and 5, respectively).  To ensure numerical stability, the
time-step on which the equations were evolved was constrained to be
\begin{equation}
dt=(dt_{\rm visc}^{-2}+dt_{\rm cour}^{-2})^{-1/2},
\end{equation}
where $dt_{\rm cour}$ is the usual CFL timestep and
\begin{equation}
dt_{\rm visc}={\cal C}_{\rm visc} \min\left[\frac{dx_i^2}{\mu}\right]
\end{equation}
is the relevant viscous timestep.  In this last expression, $dx_i$ is
the size of a grid cell in the $i$-th direction, and the minimization
occurs over all grid cells. On the basis of a standard stability
analysis, we choose ${\cal C}_{\rm visc}=0.1$.

\begin{table}
\begin{center}
\begin{tabular}{lcccc}\hline
Run & $\mu/10^{-3}$ & Re & Simulation grid \\\hline
1   & 0 & $\sim 2\times 10^3$ & $200^3$ \\
2   & 1 & 500 & $200^3$  \\
3   & 2 & 250& $200^3$  \\
4   & 4 & 125& $200^3$  \\
5   & 10 & 50 & $200^3$  \\
6   & 20 & 25 & $200^3$  \\\hline
\end{tabular}
\end{center}
\caption{Basic information for the set of simulations
presented in this paper.}
\end{table}

In real systems, the interior of the ICM cavities is filled with
extremely tenuous and strongly magnetized relativistic plasma; the
coefficient of viscosity of this material is likely to be negligible.
However, our (one-fluid) simulations model these structures simply as
bubbles of very hot gas (with initial conditions of $T_{\rm
  bubble}=100T_{\rm ambient}$).  If we were to include the temperature
dependence of $\mu$ in our calculations (eqn.~\ref{eq:mu}), the
coefficient of viscosity inside the bubble would be three orders of
magnitude greater than in the ambient ICM.  This is unphysical.  We
address this problem by fixing $\mu$ to be constant in both space and
time throughout a given simulation.  This choice suppresses the
viscosity of the simulated hot gas inside the bubble.  We note that
this method avoids the need to artificially truncate the viscosity
within the bubble (e.g., see Ruszkowski, Br\"uggen \& Begelman 2004),
leading to greater robustness and numerical stability.  We have,
however, performed additional simulations in which we do truncate the
viscosity within the bubble (using a density threshold) in order to
assess the impact of this assumption.  Comparing those simulations
with the ones presented in this paper, we confirmed that the dynamics
of the buoyant bubble considered here are not qualitatively effected
by the constant $\mu$ assumption.

Using the same definition as Section~\ref{sec:viscosity}, the Reynolds
number characterizing the hydrodynamics of the simulated bubble is
\begin{equation}
Re\equiv \frac{vL\rho}{\mu}=50\left(\frac{v}{2c_s}\right)\left(\frac{L}{1\,{\rm unit}}\right)\left(\frac{\rho}{1\,{\rm unit}}\right)\left(\frac{\mu}{10^{-2}}\right)^{-1},
\end{equation}
where $v$ and $L$ are the characteristic velocity and size,
respectively, of the bubble.  We perform a set of simulations to
explore a range of Reynolds numbers; see Table~1 for details of the
runs performed.

\subsection{The inviscid (control) case}

\begin{figure}
\hbox{
\psfig{figure=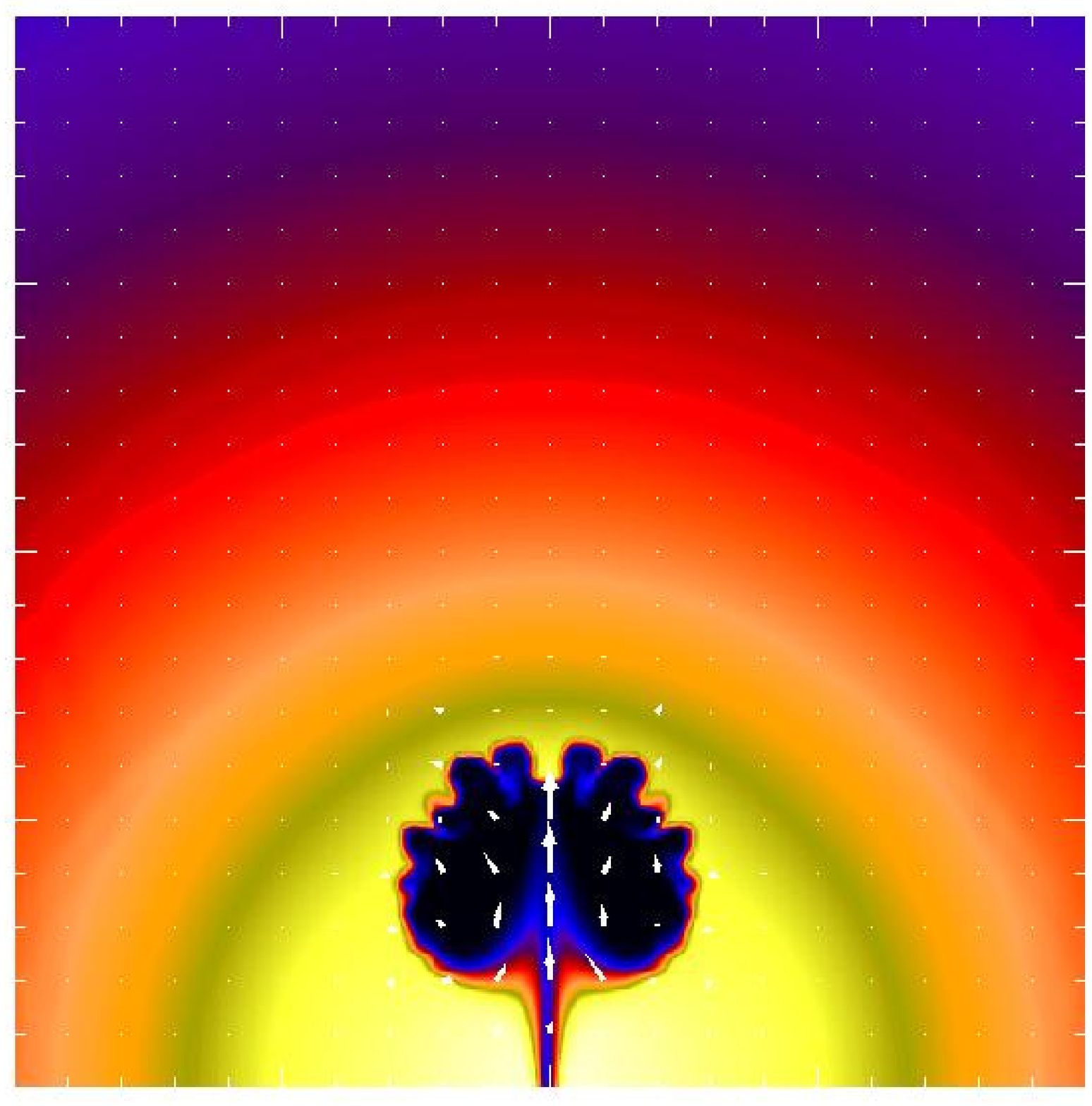,width=0.30\textwidth}
\hspace{0.03\textwidth}
\psfig{figure=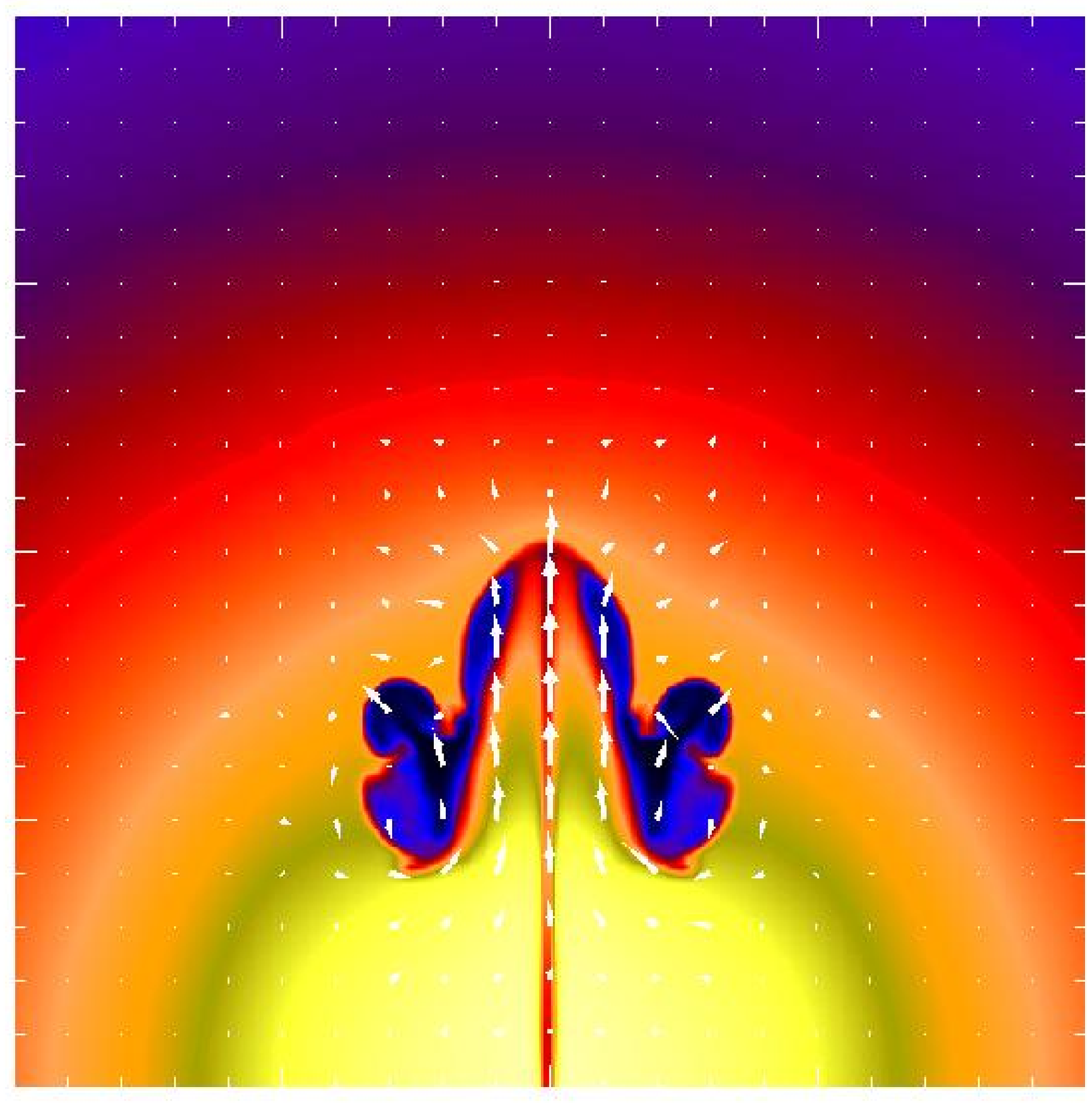,width=0.30\textwidth}
\hspace{0.03\textwidth}
\psfig{figure=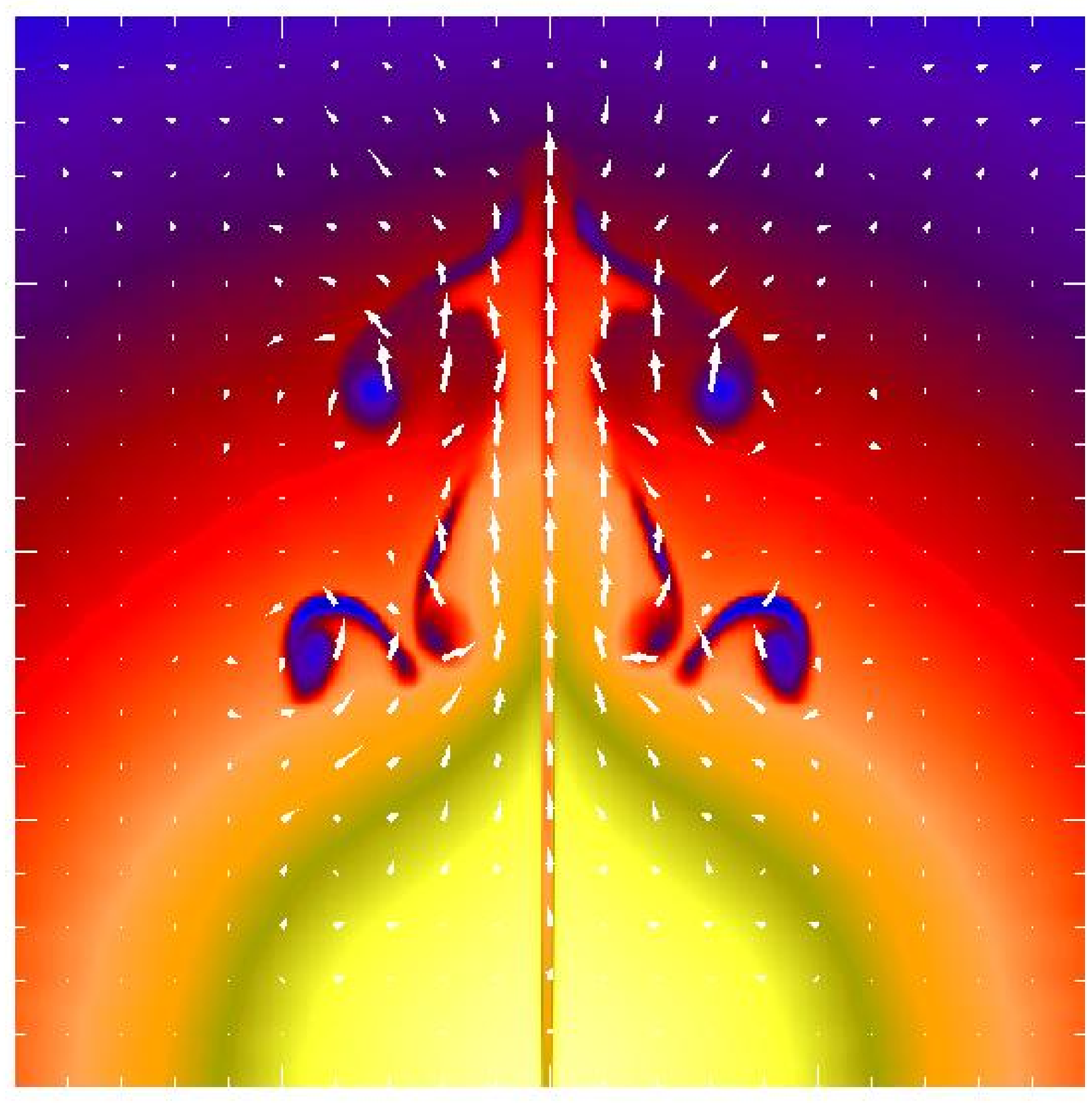,width=0.30\textwidth}
\hspace{0.03\textwidth}
}
\vspace{0.5cm}
\hbox{
\psfig{figure=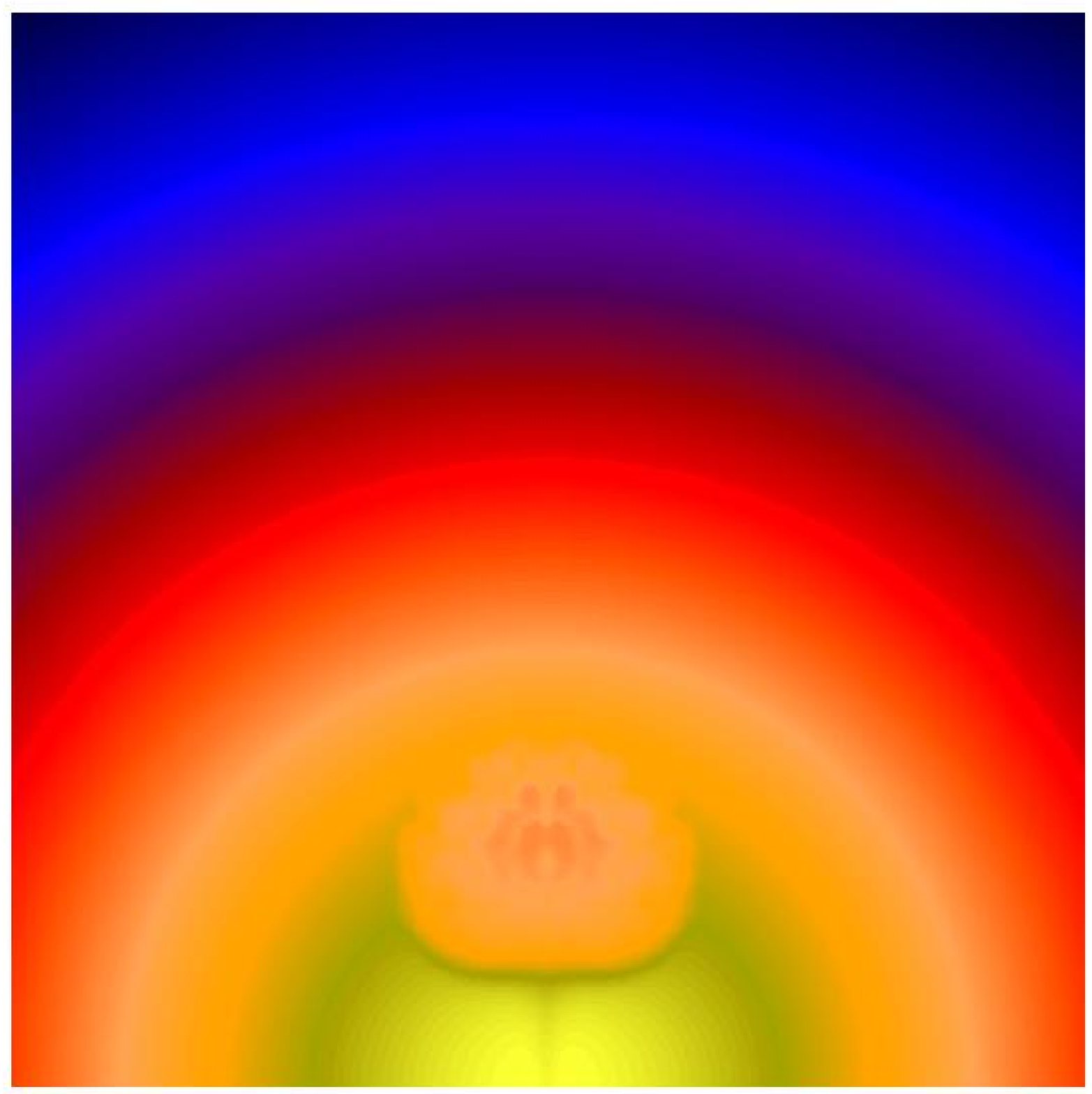,width=0.30\textwidth}
\hspace{0.03\textwidth}
\psfig{figure=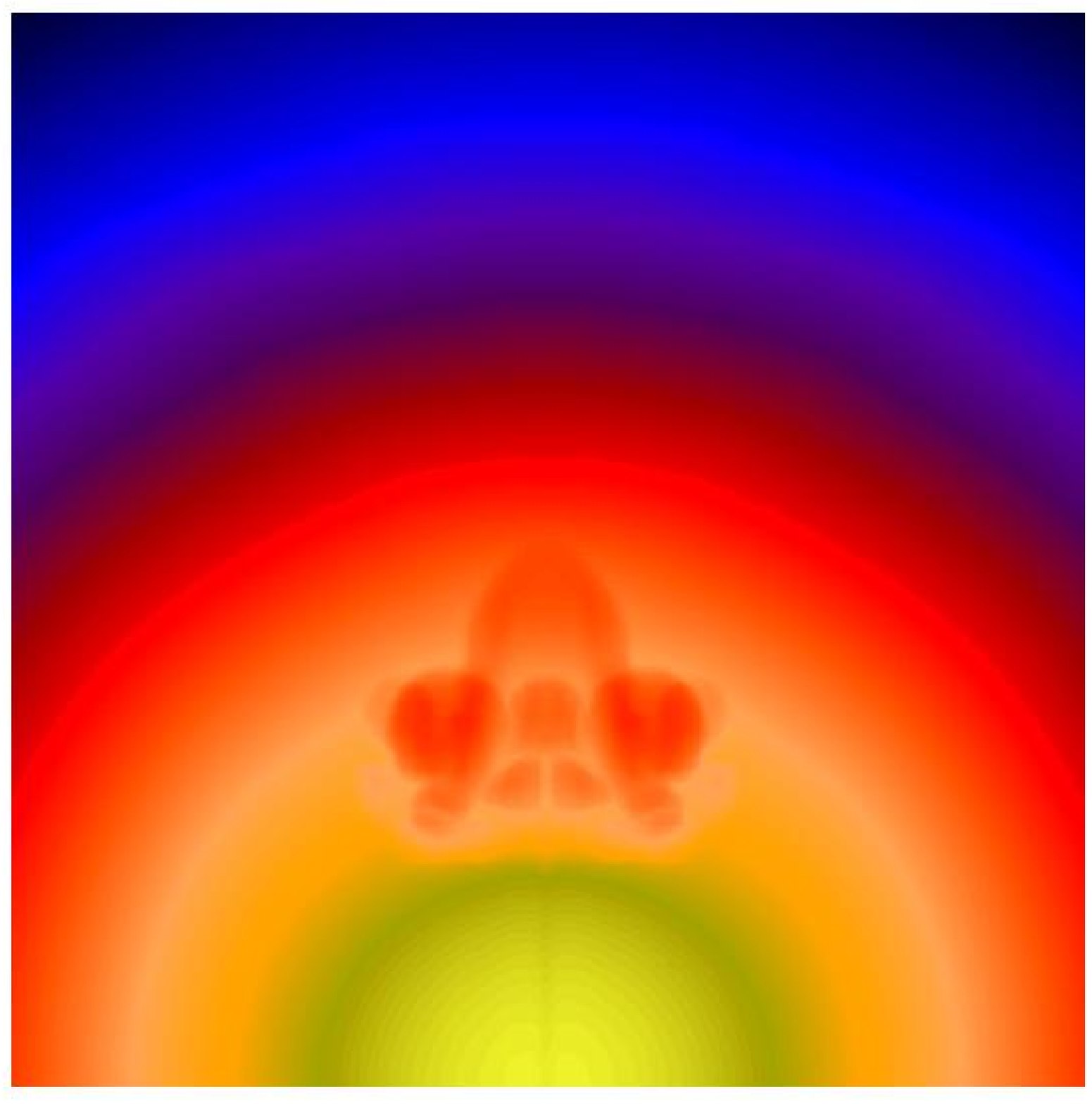,width=0.30\textwidth}
\hspace{0.03\textwidth}
\psfig{figure=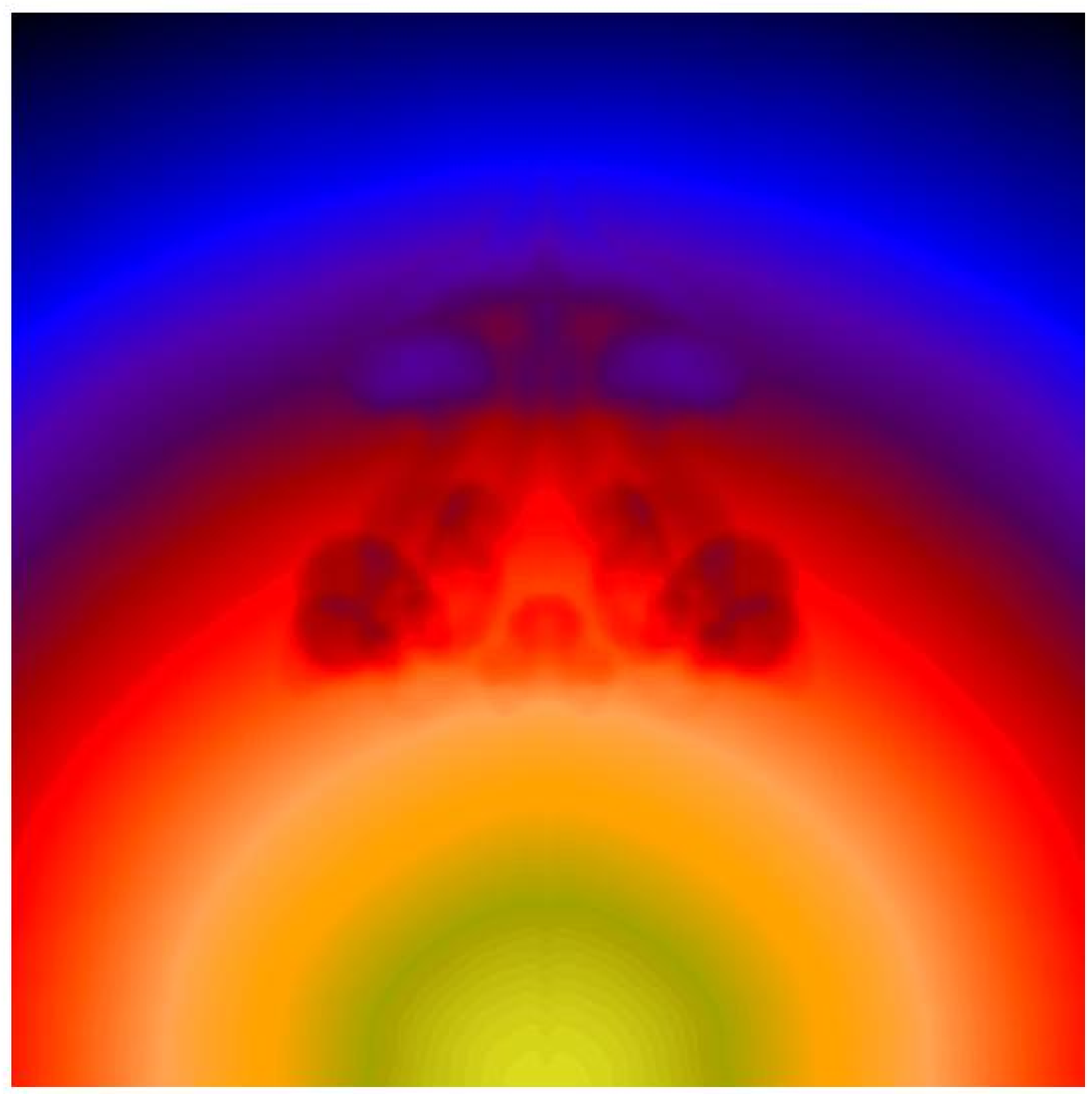,width=0.30\textwidth}
\hspace{0.03\textwidth}
}
\caption{Mid-plane density slices (upper panels) and simulated X-ray surface 
  brightness maps (lower-panels) for the inviscid control case (Run~1)
  shown at three times; $t=1$ (left panels), $t=2$ (middle panels) and
  $t=4$ (right panels).  Arrows indicating fluid velocity have been
  superposed on the density slices.  Note how the bubble is rapidly
  destroyed by the combined action of RT and KH instabilities.  At no
  time would one observe a flattened detached ghost cavity as we
  see in Per-A.}
\label{fig:novisc}
\end{figure}

Initially we discuss the results from our zero-viscosity ($\mu=0$)
simulations.  These will serve as a crucial comparison for
understanding the viscous simulations that we shall discuss next.  Our
qualitative simulation setup and results are very similar to those
previously obtained by Churazov et al. (2001), Br\"uggen \& Kaiser
(2001, 2002), Br\"uggen et al. (2002) and Robinson et al.  (2003).  Of
course, it is important to realize that even in this $\mu=0$ cases,
the action of numerical diffusion keeps the effective Reynolds number
finite.  We determine the effective Reynolds number of these
``zero-viscosity'' cases by performing additional simulations with
small values of $\mu$ and visually comparing the smallest scale
structures resulting from the $\mu=0$ Run.  This exercise suggests
that the effective Reynolds numbers of our $\mu=0$ simulations are in
the range 2000--5000.

The initially static bubble starts accelerating due to buoyancy.  As
noted by several previous authors, RT instabilities induce circulatory
motions within the bubbles, that then induce ``secondary'' KH
instabilities along the contact discontinuity between the low-density
bubble and ambient ICM.  These KH instabilities are primarily
responsible for shredding the bubble within 2--3 time units (i.e.
$\sim 5$ sound crossing times of the bubble; top-panels of
Fig.~\ref{fig:novisc}).  Due to the shredding of the bubble, one never
observes a detached and flattened but otherwise intact structure such
as we appear to see in the ghost cavity of Perseus-A (see bottom
panels of Fig.~\ref{fig:novisc}).

To further facilitate comparison with observations, we produce
simulated X-ray surface brightness maps.  In detail, we set the local
X-ray emissivity to be proportional to $\rho^2 T^{1/2}$ and integrate
along lines of sight through the simulation volume in order to build
up a 2-dimensional map of X-ray surface brightness.  For definiteness,
we show surface brightness maps for the case in which the observer is
viewing along the $y$-direction (i.e., the line joining the center of
the ICM atmosphere and the center of the initial bubble is
perpendicular to the observers line of sight).  From these maps, it
can be seen that the observed cavity never has the appearance of the
Perseus-A ghost cavity (Fig.~\ref{fig:novisc}; lower panels).

In addition to morphological problems, we note that zero-viscosity
models does not reproduce the H$\alpha$-deduced flow pattern.  At
early times, while the buoyant bubble is still reasonably intact, we
never see a circulatory flow pattern below the bubble.  At late times,
the buoyant bubble loses integrity and induces complex and
disorganized motions in the ICM.

\subsection{Simulations including viscosity}

\begin{figure}
\hbox{
\psfig{figure=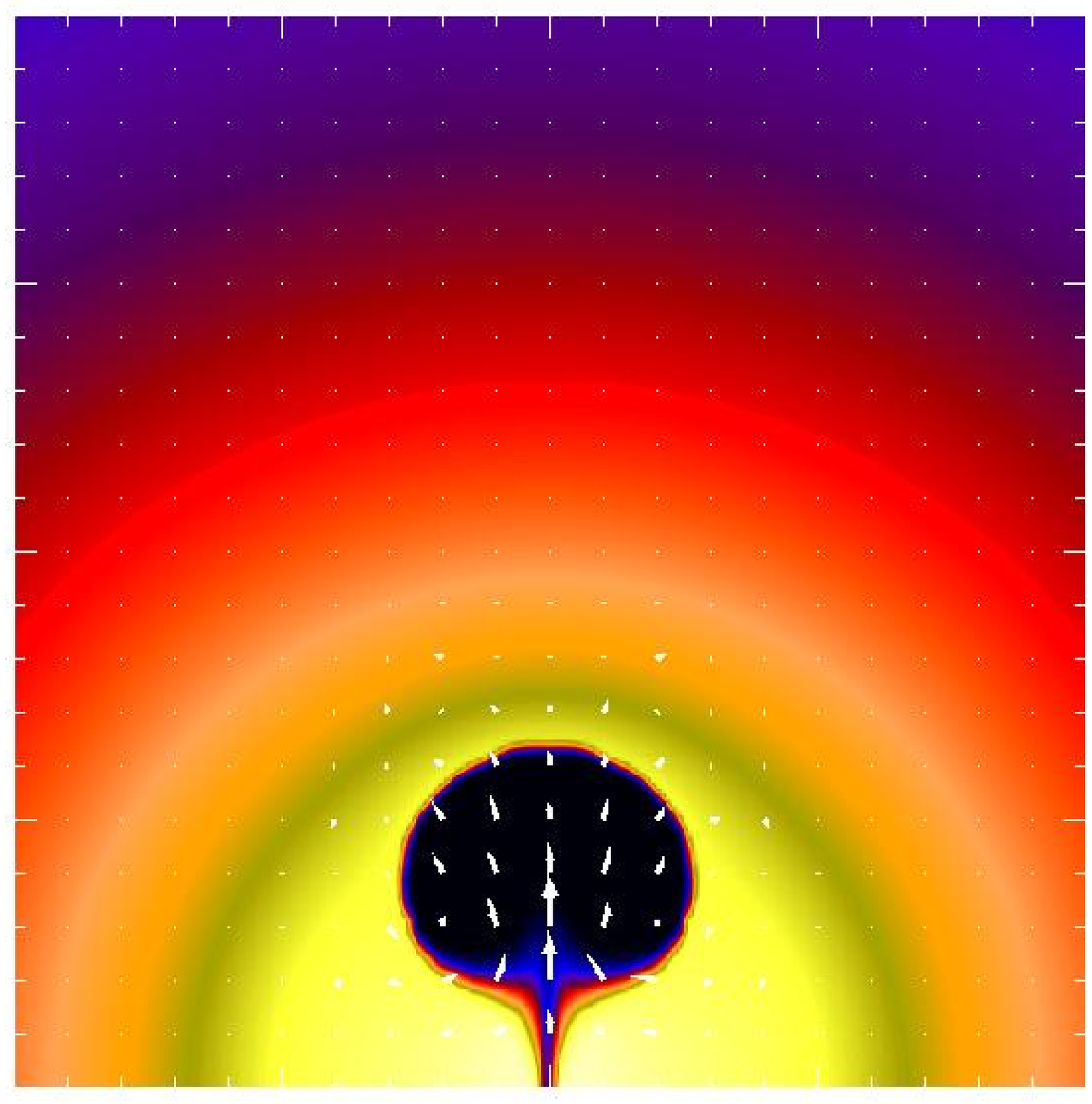,width=0.30\textwidth}
\hspace{0.03\textwidth}
\psfig{figure=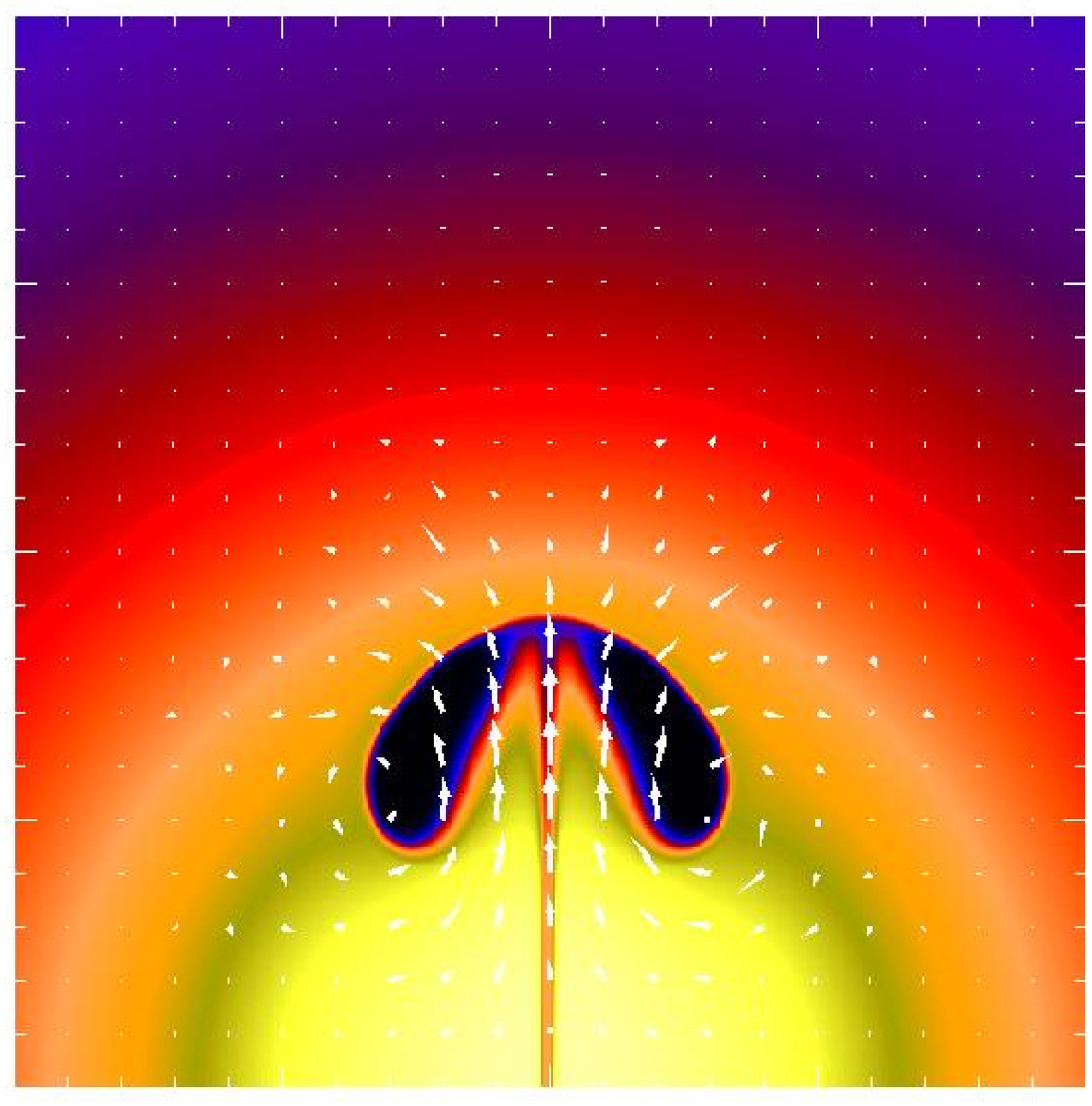,width=0.30\textwidth}
\hspace{0.03\textwidth}
\psfig{figure=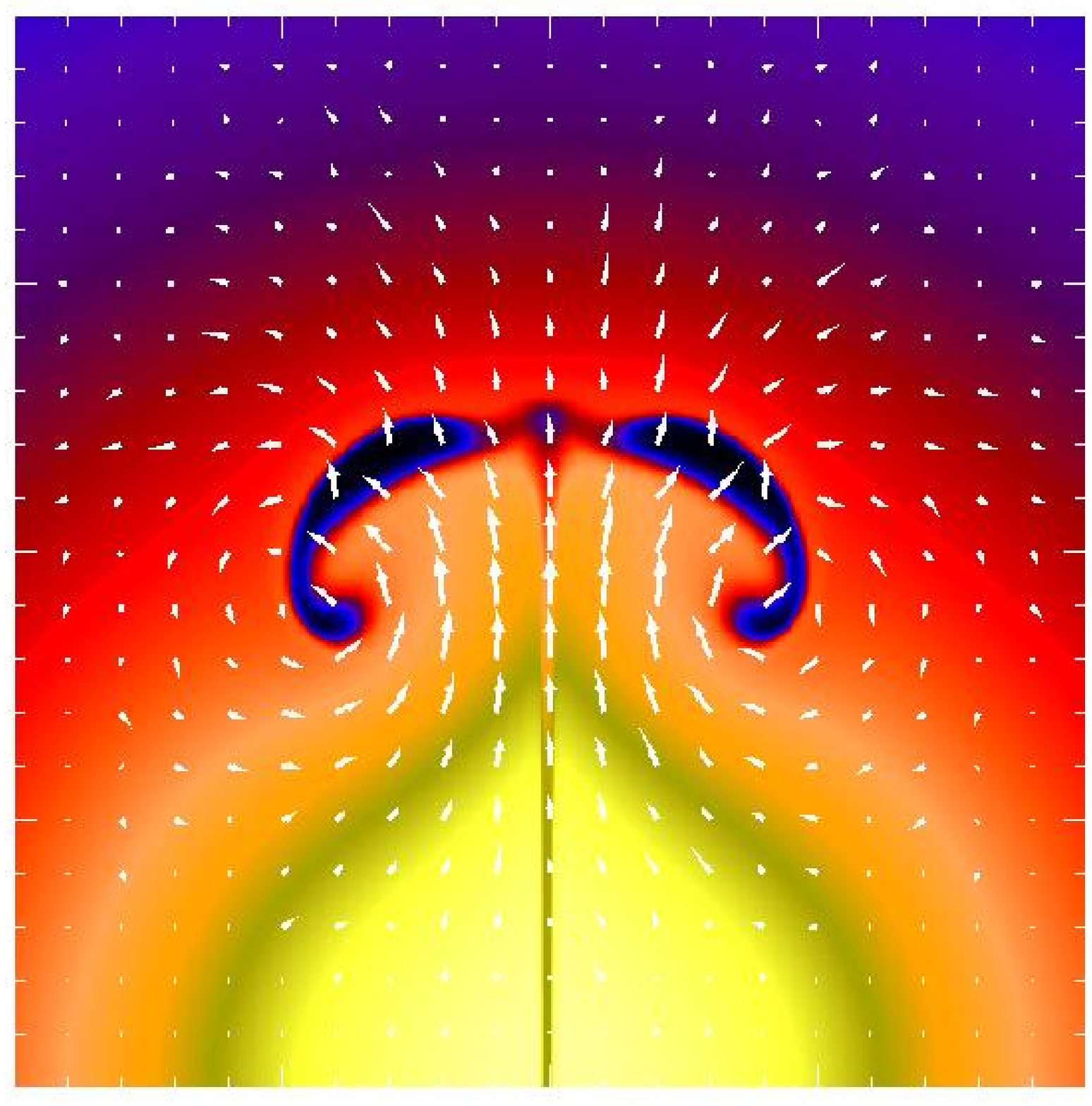,width=0.30\textwidth}
\hspace{0.03\textwidth}
}
\vspace{0.5cm}
\hbox{
\psfig{figure=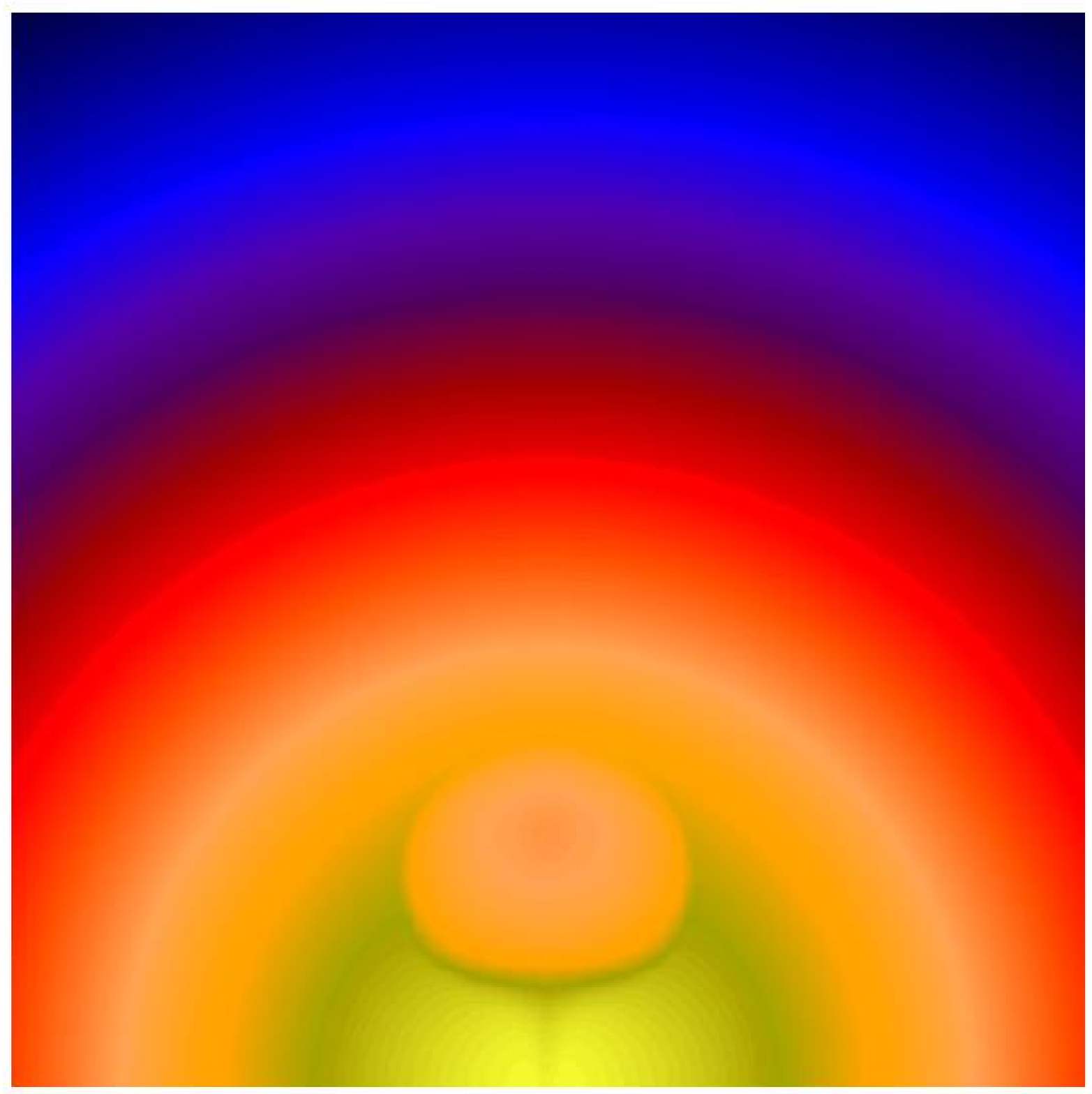,width=0.30\textwidth}
\hspace{0.03\textwidth}
\psfig{figure=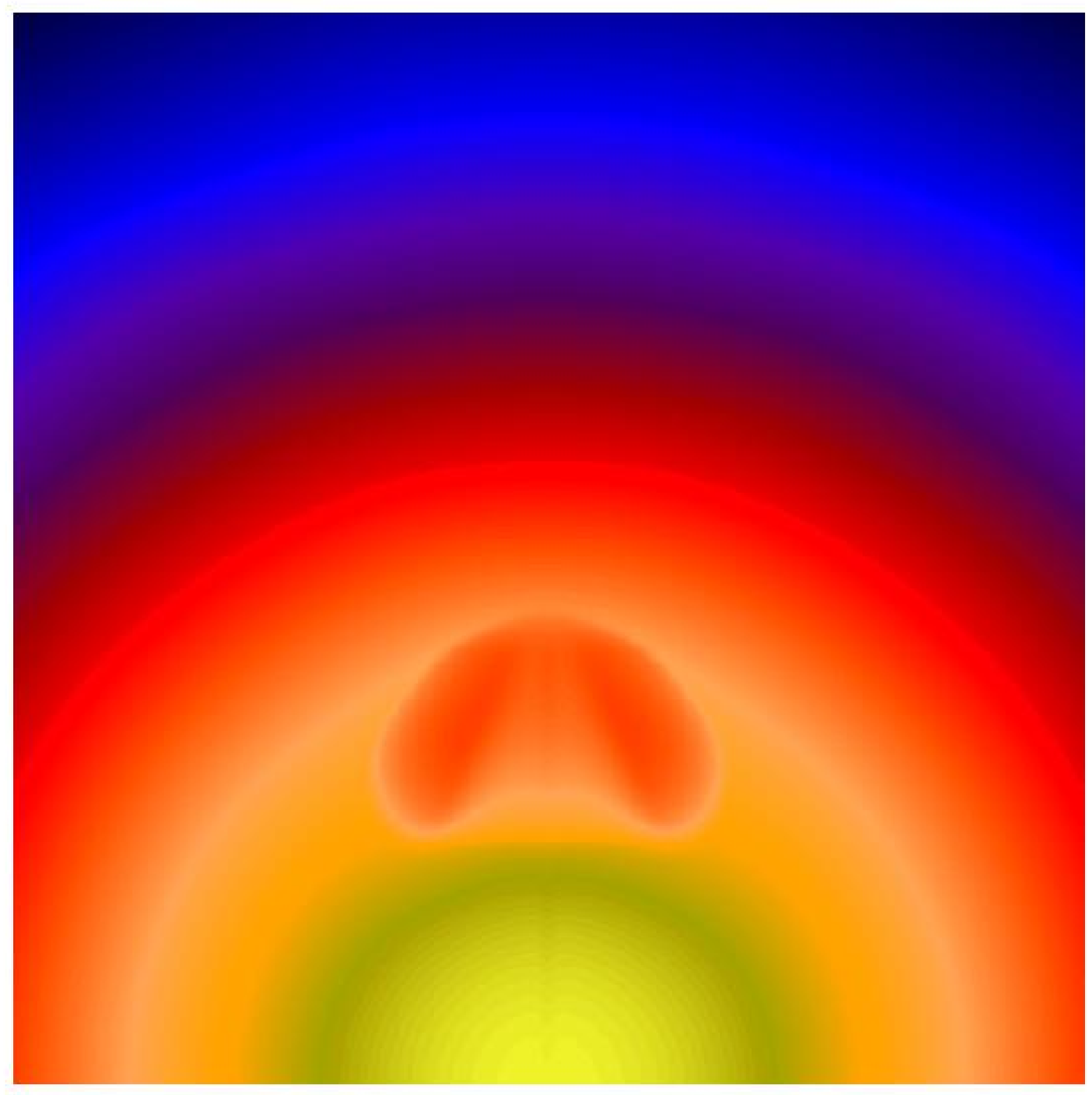,width=0.30\textwidth}
\hspace{0.03\textwidth}
\psfig{figure=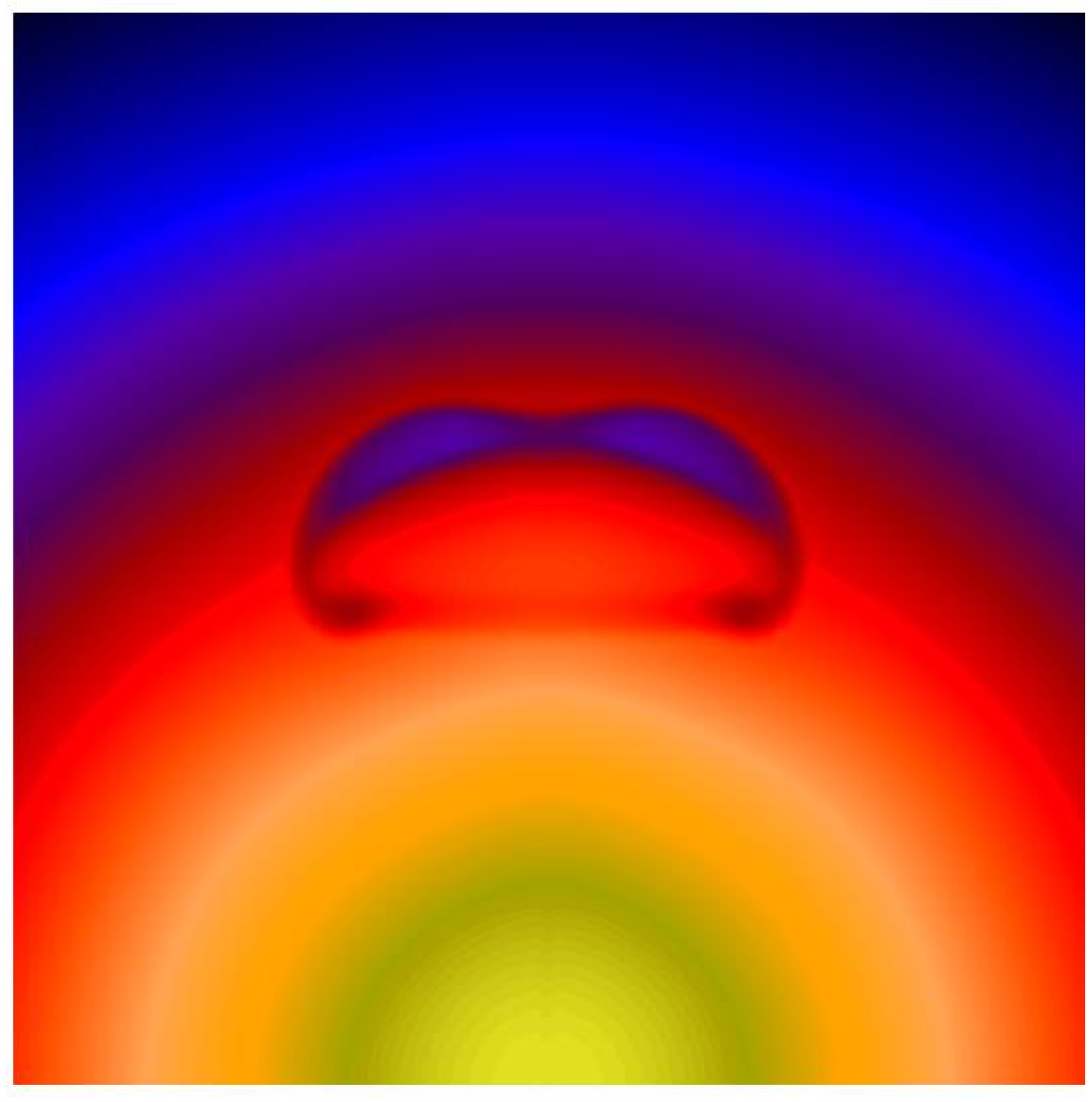,width=0.30\textwidth}
\hspace{0.03\textwidth}
}
\caption{Mid-plane density slices (upper panels) and simulated X-ray surface 
  brightness maps (lower-panels) for the $Re=250$ case (Run~3) shown
  at three times; $t=1$ (left panels), $t=2$ (middle panels) and $t=4$
  (right panels).  Arrows indicating fluid velocity have been
  superposed on the density slices. Note how the viscosity stabilizes
  the bubble, allowing a flattened by intact buoyant ``cap'' to form.
  Both the X-ray surface brightness and H$\alpha$-inferred velocity
  field around the ghost cavity of Per-A can be qualitatively
  reproduced by this model.}
\label{fig:mu002}
\end{figure}

Having described the inviscid ``control'' case, we now proceed to
discuss the effect of viscosity on the buoyant evolution of
radio-lobes.  As in the inviscid case, the evolution is driven by the
joint action of buoyancy and secondary KH instabilities.  However,
unlike the inviscid case where KT instabilities operate at the contact
discontinuity on spatial scales down to the grid scale, viscosity
suppresses the KH instability on small spatial scales.  This has a
profound effect on the evolution of the bubble; even a moderate amount
of viscosity can prevent the shredding of the bubble, which can
subsequently float out of the core being rather flattened but
otherwise intact.  

As a specific example, Fig.~\ref{fig:mu002} shows the $Re=250$
(Run~3).  This can be considered a model of the ghost cavities around
Perseus-A if the ICM possesses a coefficient of viscosity of
$\mu\approx 0.25\mu_{\rm um}$.  As discussed in
Section~\ref{sec:viscosity}, this level of viscosity may be plausible
even in the presence of tangled, chaotic magnetic fields.  It can be
seen from the mid-plane density and velocity fields
(Fig.~\ref{fig:mu002}; upper panels) that the evolution of the bubble
is driven by buoyancy, with secondary KH instabilities largely unable
to overcome the action of the viscosity.  As the bubble floats
upwards, it flattens into a broad cap.  The surface brightness maps
associated with Run~3 (Fig.~\ref{fig:mu002}; lower panels) show that
one does, indeed, produce a detached and flattened cavity in the ICM
emission as observed in the Perseus cluster.

\begin{figure}
\centerline{
\psfig{figure=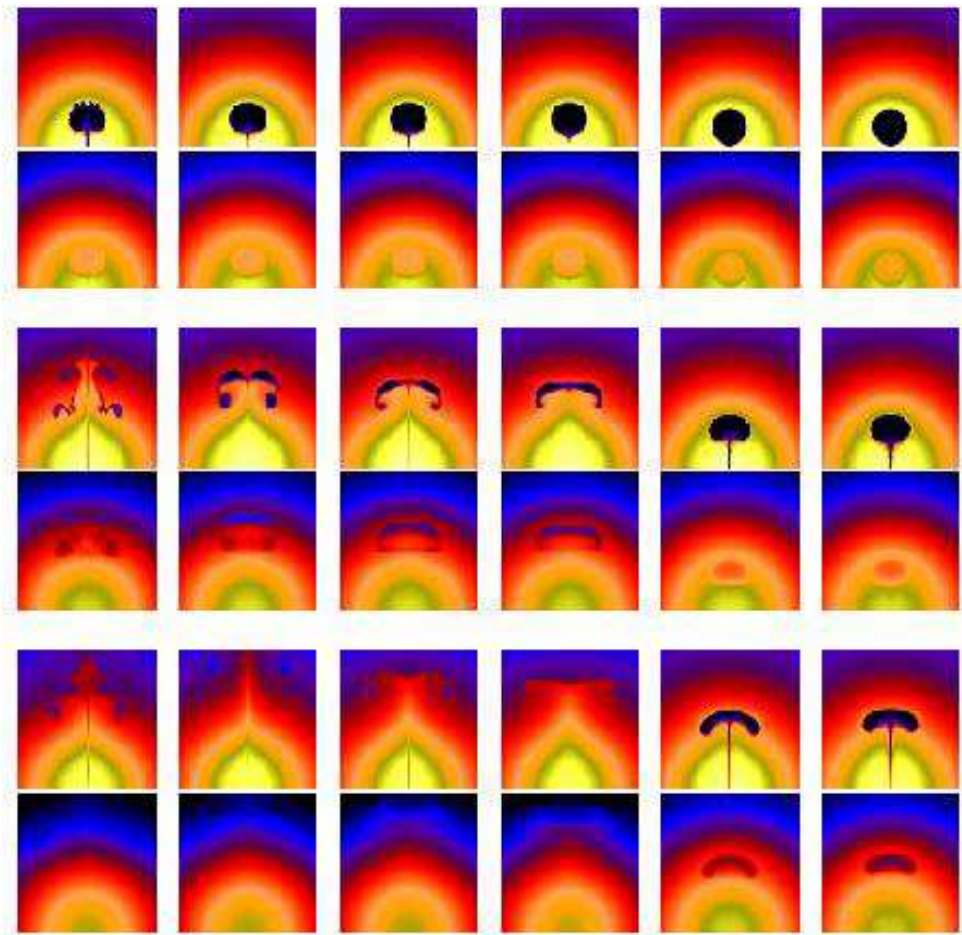,width=1.0\textwidth}
}
\caption{Mid-plane density slices (upper panels of each set) and 
  simulated X-ray surface brightness maps (lower-panels of each set)
  for the six simulations presented in this paper (Runs 1--6 ordered
  from left to right).  Results are shown for three times, $t=1$
  (upper set of panels), $t=4$ (middle set of panels) and $t=8$ (lower
  set of panels).  See text for a discussion of these results.}
\label{fig:visc_range}
\end{figure}

Figure~\ref{fig:visc_range} shows results for the full range of
viscosity explored in this paper at three fixed times ($t=1, 4, 8$).
It can be seen that the formation of an flattened but intact buoyant
bubble occurs in all of our viscous simulations.  However, the
timescale on which the evolution proceeds is a strong function of the
viscosity.  For example, the $t=1$ density slice of Run~2 ($Re=500$)
is very similar to the $t=4$ slice of Run~6 ($Re=50$).  This fact may
point to a solution of the ``shock problem'' noted in the
introduction, an issue that we shall return to in
Section~\ref{sec:shock}.

Viscosity also has important implications for the flow pattern in the
disturbed ICM.  In principle, the presence of viscosity can facilitate
the development of large scale vortex rings in the trailing region
beneath the rising bubble.  This phenomena is seen for our highest
viscosity cases.  For the levels of viscosity that are probably
relevant to the Perseus cluster (${\rm Re}=100-200$), this effect is
not seen.  However, even for these levels of viscosity, our
simulations show flow patterns that qualitatively match those inferred
from the H$\alpha$-filament geometry in Perseus.  In these cases, the
flattened buoyant bubble undergoes a minor fragmentation due to the
action of secondary KH instabilities.  As a result of this
fragmentation, a small torus of radio plasma is left in the trailing
region behind the main buoyant bubble.  The simulations show a strong
ICM circulation around this trailing torus, producing a streamlines
that resemble the H$\alpha$ filament geometry in Perseus.  A
prediction of this model is that sufficiently sensitive X-ray maps
should reveal subtle depressions at the center of these vortices, and
sufficiently sensitive and spatially resolved low-frequency radio maps
should reveal a corresponding torus of aged radio plasma.

\begin{figure}
\centerline{
\psfig{figure=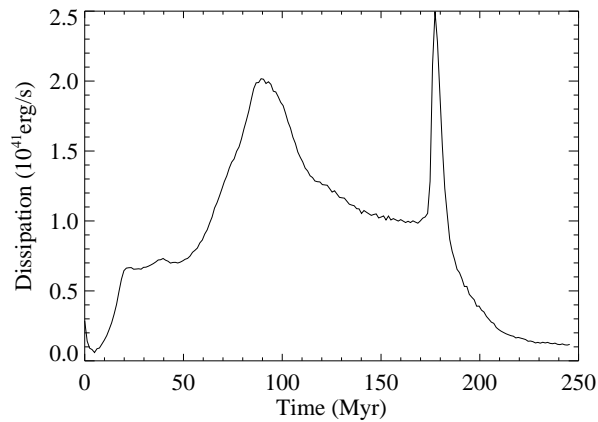,width=0.49\textwidth}
}
\caption{Volume integrated viscous dissipation rate for our $Re=250$ case, 
  scaled with parameters relevant for the Perseus cluster (i.e., the
  simulation domain has a linear size of 40\,kpc in each dimension,
  one code unit of density is $0.03\pcmcu$, and the initial ICM sound
  speed is $780\kmps$).  One code unit of time is then 24\,Myr.  As
  noted in the main text, the second and rather sharp peak in the
  dissipation rate is likely due to an interaction of the flow pattern
  with the boundaries of the computation and hence should not be
  considered physical.}
\label{fig:dissipation}
\end{figure}

An important consequence of ICM viscosity is that it provides an
explicit mechanism by which the radio-galaxy induced disturbance can
heat the ICM.  While a full treatment of viscous ICM heating by
radio-galaxies almost certainly requires following the jet-driven
inflation of the bubbles (e.g., Reynolds, Heinz \& Begelman 2002), as
well as multiple epochs of activity (Ruzskowski, Br\"uggen \& Begelman
2004), it is instructive to compute the viscous dissipation rate in
our idealized simulation.  Figure~\ref{fig:dissipation} shows the
dissipation rate as a function of time for our canonical
``Perseus-like'' model, the $Re=250$ case.  The dissipation rate and
the time coordinate are given in physical units assuming parameters
relevant to the Perseus cluster (see caption of
Fig.~\ref{fig:dissipation}).  The dissipation rate dispays two peaks.
The rather broad peak centered at about 90\,Myr coincides with the
secondary KH instability entering the strongly non-linear regime, and
the subsequent ``folding'' of the flattened bubble.  The second and
rather sharp peak corresponds to the venting of material out of the
boundary of the simulation and, hence, should not be considered
physical.  In general, the dissipated power achieves the rather modest
levels of $P\sim 10^{41}\ergps$.  However, this heat source continues
to operate for a period of about 200\,Myr, an order of magnitude more
than the plausible recurrence timescale of the radio-galaxy activity
(Fabian et al.  2003a).  Thus, the possibility of balancing the
radiative losses with viscous dissipation from the combined effect of
many bubbles remains open.  Furthermore, it is likely that the
dissipation of the fluid modes driven by the initial inflation of the
bubble (and hence not modelled here) will deliver just as much energy,
if not more, than the viscous dissipation during the buoyant phase.  

\section{Discussion}

\subsection{The neglect of a jet and magnetic fields}

Our simulations must be viewed as a rather limited toy model for the
buoyant evolution of radio-lobes and the formation of ghost cavities.
We have employed two major simplifications.  Firstly, we have not
modeled the jet-driven inflation of the bubble.  Real examples of
these radio-lobes will possess complex, jet-driven flows with speeds
much in excess of the ICM sound speed, and will never resemble the
static bubbles that we use for our initial conditions (Reynolds, Heinz
\& Begelman 2002).  Consequently, the KH instability will be important
at all stages in the life of the bubble, and does not need to wait for
the RT instability to first induce bulk flows.

Secondly, we have neglected the effects of magnetic field.  It is
expected that fields within the radio plasma could readily achieve
equipartition strengths and hence have an important effect on the
dynamics of the rising bubble.  In particular, magnetic fields will
influence the development of both the RT and the KH instability at the
contact discontinuity.  The effect of magnetic fields on the buoyant
evolution of ICM bubbles has recently been addressed in the context of
inviscid hydrodynamics by Br\"uggen \& Kaiser (2001), Robinson et al.
(2003) and De Young (2003).  Their results suggest that magnetic
fields can have a strongly stabilizing effect on buoyant bubbles,
effectively quenching all of the instabilities that distort and
eventually shred this structures.  However, both the symmetry of their
calculation (2-d Cartesian) and their initial magnetic field
configuration (taken from Cargill et al. 1996) makes the Robinson et
al. (2003) calculation most relevant to a buoyantly rising {\it
  cylindrical flux tube with purely toroidal magnetic flux}, a poor
approximation to an AGN blown bubble.  The ability of magnetic fields
to stabilize or de-stabilize the contact discontinuity is a strong
function of the magnetic configuration and field strength (e.g., see
Jun, Norman \& Stone 1995), both of which are open issues in the case
of AGN-blown bubbles.

Clearly further simulations are required to address these two issues.
In fact, it would be most appropriate to treat these as coupled
questions.  It is believed that the magnetic field in such bubbles
originates from the magnetized AGN jets.  Thus, we must follow the
jet-driven inflation of the bubble in order to produce a reasonable
initial field configuration.  This will be the subject a future
publication.

\subsection{Can viscosity solve the ``shock problem''?}
\label{sec:shock}

In addition to stabilizing buoyant radio-lobes against shredding, we
have discussed how viscosity can dramatically slow the evolution of
the bubble.  This suggests a natural solution to the ``shock problem''
noted in the introduction.  Under the action of viscosity, the AGN
jets may inflate their ICM bubbles subsonically before the buoyancy
effects become important.  Consequently, the ICM can be displaced into
a cavity-bounding shell without the need of a strong shock.  The
thermal structure of this gas will be determined by the action of
compression (due to its displacement from the cavity), decompression
(due to the fact that this material will typically be lifted to a
higher level in the ICM atmosphere), viscous dissipation, and
radiative cooling.  One can envisage a scenario in which the rather
slow evolution of the radio-lobe in a viscous ICM allows the shell of
displaced ICM to radiatively cool by an appreciable amount.  Further
simulations of the inflation stage of these bubbles are required to
assess whether the rather cool cavity-bounding shells observed in
Per-A can be reproduced within the context of this model.  Given the
strong temperature dependence of the coefficient of viscosity
(eqn.~1), we might expect significant changes in the effective
viscosity between the various ICM structures defining the system
(i.e., the ambient material, the cool rims and the shocks, if any).
Thus, it will be important to move beyond the constant $\mu$
assumption in these next generation of simulations.

\section{Conclusions}

The true role of kinematic viscosity in ICM/radio-galaxy interactions
remains unclear, primarily due to the great uncertainty associated
with magnetic suppression of Spitzer-type transport processes in the
plasma.  However, in stark contrast to the inviscid models, our models
of bubble evolution in modestly viscous ICM can reproduce both the
morphology and inferred flow patterns seen around the best studied of
the ghost ICM cavities, the NW cavity in Perseus-A.  The principal
effect of viscosity is to stabilize the bubble against both RT and KH
instabilities, thereby allowing it to remain intact as it floats
upwards in the ICM atmosphere.  The flattening of such bubbles as they
rise is a natural explanation for the elongated morphology of the NW
cavity in Perseus-A.  The ``smoke-ring'' like flow pattern below the
NW cavity inferred from the morphology of the H$\alpha$ filaments can
then arise due to minor fragmentation of the rising bubble (leading to
a trailing torus) or genuine vortex shedding.

If the ICM is indeed viscous at the level treated by our models, the
subsequent slowing of the evolution timescales for ICM/radio-galaxy
interactions may be the solution to the ``shock problem'', i.e. the
lack of strong shocks bounding the ICM cavities that are associated
with active radio-lobes.  In this scenario, radio galaxies such as
Per-A would be rather less powerful than previously thought (Heinz,
Reynolds \& Begelman 1998), inflating their associated ICM bubbles
subsonically.  If the evolution is sufficiently slow, radiative
cooling might allow the ICM shell surrounding the cavity to cool,
again in agreement with Chandra observations of Perseus-A.

Ultimately, disentangling the effects of transport processes
(viscosity and thermal conduction) and magnetic fields will require
further simulation work coupled with detailed observations.  On the
theoretical side, it is important to initiate a new generation of
simulations which incorporate full MHD {\it as well as} transport
processes.  Qualitatively new and rich dynamics becomes possible upon
the inclusion of transport processes within MHD.  For example, Balbus
(2000) discovered the ``magnetothermal instability'' which occurs in
any MHD atmosphere with a temperature profile that decreases with
height provided there is thermal conduction along the magnetic field
lines. In addition, the viscous transport coefficients will be
anisotropic in the presence of a magnetic field, and this may lead to
yet further complexity in the dynamics.  Such instabilities could well
be crucial for understanding the dynamics of radio-galaxy/ICM
interactions.  Observationally, deep (megasecond) {\it Chandra}
observations of the nearest radio-galaxy/cluster interactions might
allow detailed imaging spectroscopy of fluid instabilities associated
with the radio-galaxy/ICM interaction.  Furthermore, upon the launch
of {\it Astro-E2} in 2005, we will be able to directly probe
turbulent velocities within the ICM of nearby (bright) clusters
through broadening of their X-ray emission lines.  With the caveat
that full MHD viscous simulations have yet to be done, the presence or
absence of turbulence in the ICM, especially in regions well separated
from potential driving sources (such as AGN or merging subclusters)
could be an important indicator of the presence of viscosity.

\section*{Acknowledgments}

We thank Cole Miller and Eve Ostriker for stimulating conversations
conducted throughout the course of this work.  We gratefully
acknowledge support from the National Science Foundation under grant
AST0205990 (CSR), the {\it Chandra} Cycle-5 Theory \& Modelling
program under grant TM4-5007X, and the Royal Society (ACF).  All
simulations presented in this paper were performed on the Beowulf
cluster in the Department of Astronomy at the University of Maryland,
kindly administered by Derek~Richardson.

\label{lastpage}

\end{document}